\documentstyle[12pt,aas2pp4,twoside]{article}
\newcounter{figa}
\newcounter{figb}

\begin{document}
\parindent=2em
\slugcomment{to be published in The Astrophysical Journal, Oct. 1998}

\def\etal{et~al.~}
\lefthead{Hunsberger, Charlton, \& Zaritsky}
\righthead{Luminosity Function of HCG Galaxies}


\Large
\centerline{{\bf The Luminosity Function of Galaxies in Compact Groups}
\footnote{Observations were made on the 60 inch telescope at Palomar
Mountain, which is jointly operated by the California Institute of Technology
and the Carnegie Institution of Washington.}}

\vspace{0.25in}
\large
\centerline{Sally~D.~Hunsberger and Jane~C.~Charlton\altaffilmark{2}}
\normalsize
\affil{{\rm Astronomy \& Astrophysics Department \\
       Pennsylvania State University \\
       University Park, PA 16802 \\}
       sdh@astro.psu.edu \\
       charlton@astro.psu.edu}

\vspace{0.25in}
\and
\large
\centerline{Dennis~Zaritsky\altaffilmark{3}}
\normalsize
\affil{{\rm UCO/Lick Observatories and
       Astronomy \& Astrophysics Department, \\
       University of California, Santa Cruz, CA 95064 \\}
       dennis@ucolick.org}

\altaffiltext{2}{Center for Gravitational Physics and Geometry,
Pennsylvania State University}
\altaffiltext{3}{Alfred P. Sloan Research Fellow}

\vspace{0.25in}
\centerline{To be published in {\it The Astrophysical Journal}}

\vspace{0.25in}
\centerline{\bf ABSTRACT}
From R-band images of 39 Hickson compact groups (HCGs), we use galaxy counts
to determine a luminosity function
extending to $M_R=-14.0 + 5\log h_{75}$, approximately two
magnitudes deeper than previous compact group luminosity functions.
We find that a single Schechter function (Schechter 1976) is a poor fit
(${\chi}_{\nu}^2 > 4$) to the data,
so we fit a composite function consisting of separate Schechter functions for
the bright and faint galaxies.
The bright end is best fit with $M^*=-21.6$ and $\alpha=-0.52$ and the
faint end with $M^*=-16.1$ and $\alpha=-1.17$.
The decreasing bright end slope implies a deficit of 
intermediate luminosity galaxies
in our sample of HCGs and the faint end
slope is slightly steeper than that reported for earlier HCG luminosity functions.
Furthermore, luminosity functions of subsets of our sample reveal more 
substantial dwarf populations for groups with x-ray halos, groups with
tidal dwarf candidates, and groups with a dominant elliptical
or lenticular galaxy.
Collectively, these results support the hypothesis that within compact groups,
the initial dwarf galaxy population is 
replenished by  ``subsequent generations'' formed in the tidal debris of giant
galaxy interactions.

\vspace{0.25in}
{\it Subject headings: } galaxies --- luminosity function, compact groups, evolution


\clearpage
\pagestyle{myheadings}

\markboth{~~{\sc Hunsberger, Charlton, \& Zaritsky} \hfill Luminosity Function of HCG Galaxies}
         {{\sc Hunsberger, Charlton, \& Zaritsky} \hfill Luminosity Function of HCG Galaxies~~}
\section{Introduction}
The dwarf galaxy population is not merely an extension of giants to fainter
luminosities.
Fundamental differences between the two populations raise the issue of whether the same
formation mechanism applies to both.
Understanding the origin and evolution of dwarf galaxies is essential in
developing the correct model of galaxy and cluster formation.
Because compact groups are unique environments and active sites of galaxy interaction, they
provide an opportunity to study the formation of dwarf galaxies both as a function of environment and in tidal
debris.

Hickson (1982) cataloged 100 compact groups of galaxies,
selecting them
on the basis of population, isolation, and compactness.
Radial velocity measurements (Hickson et al. 1992) and morphological studies
(Mendes de Oliveira \& Hickson 1994) suggest that most of the Hickson compact
groups (HCGs) are
physical associations, although Mamon (1990) has argued that the data are
also consistent with many HCGs being superpositions of binary-rich loose
groups and Hernquist et al. (1995) contend that many compact groups are chance
projections of large filamentary structures.
Diaferio et al. (1994) propose that during the collapse of a rich loose group
of galaxies,
compact configurations are continuously forming and that
the members of such a compact group eventually merge, even as new galaxies are
joining the group from the surrounding region.

Regardless of the exact fraction of groups that are physical associations,
compact groups are sites of galaxy interactions and among themselves,
they provide a variety
of environments to examine (e.g., x-ray luminous vs. non-luminous, elliptical
vs. spiral dominated, tidally interacting vs. quiescent).
Compact groups are particularly intriguing because of the combination of
high galaxy projected density  (similar to the centers of rich clusters) and
low velocity dispersion (comparable to loose groups).
This combination suggests, at least for the fraction of physical groups,
that interactions and mergers occur frequently and that
during such encounters, tidal forces have sufficient time to extract significant
amounts of material, resulting in tidal tails and bridges.
As proposed by Zwicky (1956), self-gravitating objects might
develop in tidal tails and evolve into dwarf galaxies.
This idea is supported by the observation of
regions of active star formation at the ends of tidal tails in the
Superantennae and Antennae systems (Mirabel et al. 1991, Mirabel et al. 1992)
and of two dwarf galaxies in the tidal tails of Arp 105 (Duc \& Mirabel 1994).
Furthermore, numerical simulations (Barnes \& Hernquist 1992, Elmegreen et al. 1993)
confirm that gravitationally bound clumps can form in tidal tails.
Therefore, compact groups represent a unique environment in which to study the
formation of dwarf galaxies in tidal debris.
In a previous paper (Hunsberger et al. 1996), we examined a sample of 42 HCGs and
identified 47 tidal dwarf candidates in seven groups with tidal tails
and tidal arms.
We also estimated the fraction of compact group dwarf galaxies produced
by the tidal dwarf formation mechanism over the lifetime of a group
and found it to be significant ($> 30\%$).

Previously determined luminosity functions of compact groups (Heiligman \&
Turner 1980, Mendes de Oliveira \& Hickson 1991,
Sulentic \& Raba\c{c}a 1994, Ribeiro et al. 1994,
Zepf et al. 1997) have yielded
faint end slopes of $\alpha=0.0$, $\alpha=-0.2 \pm 0.9$,
$\alpha=-1.13 \pm 0.13$, $\alpha=-0.82 \pm 0.09$, and
$\alpha=-1.0$, respectively.
Some of these values are flatter than that observed in either clusters
($-1.4 < \alpha < -1.0$; Ferguson \& Sandage 1991) or
the field ($\alpha \sim -1.0$; Loveday et al. 1992, Marzke et al. 1994,
Ratcliffe et al. 1998)
implying fewer dwarfs per giant in compact groups than in the other environments.
The Heiligman \& Turner (1980) luminosity function used ten compact groups (not
necessarily HCGs) for which galaxy redshifts were available at that time.
Knowing the completeness limits of the Palomar Sky Survey from which the groups
were identified, they quantified the deficit of fainter members in their sample
by a comparison to the field luminosity function.
The number of ``missing'' galaxies was significant and
could not readily be attributed to small sample size, inaccurate photometry,
galaxy misclassification, or selection effects.
Using a sample of 68 HCGs, Mendes de Oliveira \& Hickson (1991) determined the
``best-fit'' parameters of their luminosity function with Monte Carlo
simulations of compact groups selected from various Schechter distributions.
They similarly concluded that there is a lack of low luminosity galaxies
due to environmental factors rather than selection effects and suggested that
the excess of very luminous ellipticals (the mean magnitude of HCG ellipticals
is brighter than that of Virgo cluster ellipticals) coupled with a deficit of
fainter galaxies indicates merger activity.
Sulentic \& Raba\c{c}a (1994) extended the compact group luminosity function to
fainter absolute magnitudes using the Hickson catalog but applying a correction
factor for incompleteness.
Although their adopted value for the faint end slope did not reflect the
decreasing number of low luminosity galaxies of previous luminosity functions,
they noted that a single Schechter function did not provide a good fit to both
bright and faint ends simultaneously. 
Ribeiro et al. (1994) obtained CCD images of 22 HCGs and used a
statistical method to create a luminosity function which 
probed the galaxy population fainter than the original Hickson catalog, however,
the resultant compact group luminosity function did not reveal the depletion
of faint galaxies suggested by earlier analyses.
Zepf et al. (1997) confirmed the validity of this photometric approach by
obtaining spectroscopic redshifts and producing a similar luminosity function.

Most luminosity functions are presented as
averaged functions and do not address questions
arising from environmental differences and tidal dwarf formation.
In this paper, we present luminosity functions determined by a
statistical technique, similar to that used by Ribeiro et al. (1994),
to address issues regarding dwarf galaxy formation.
The technique involves using galaxy counts in the outer regions of images to 
statistically subtract
the background/foreground contribution.
Such an approach bypasses the need to obtain
redshifts of all faint galaxies within the compact group field; so
only moderately deep imaging is required.
Using galaxy counts within 17 HCGs, the Zepf et al. (1997) luminosity function
extends down to $M_B=-14.5 + 5\log h$.
Our luminosity function uses a sample of 39 HCGs and extends to
$M_R=-13.4 + 5\log h$ ($M_R=-14.0 + 5\log h_{75}$).
Therefore, our limit is two magnitudes fainter assuming a typical dwarf galaxy
color of $B-R=1.0$,
which is the median value for Local Group dwarf galaxies
\footnote{From NED, the NASA/IPAC Extragalactic Database is operated by the
Jet Propulsion Laboratory, California Institute of Technology, under contract
with the National Aeronautics and Space Administration.}.
We compare luminosity functions of various sub-samples of HCGs
such as  spiral-dominant vs. elliptical-dominant, x-ray-rich
vs. x-ray-poor, and groups with tidal dwarfs vs. groups without tidal dwarfs
to discover how processes such as galaxy destruction,
galaxy creation, and tidal stripping modify the luminosity
function of a compact group during its evolution.

In the next section (\S2), we describe the observational procedures and
data analysis.
The method for determining a luminosity function is detailed in \S3 and
we present the results in \S4.
In \S5 the major results are summarized and discussed
and we present some
speculations as to the origin of the luminosity function differences.


\section{Data Acquisition and Analysis}
We selected compact groups from the Hickson catalog (1982)
on the basis of their angular size and the apparent magnitudes
of member galaxies (Hickson 1982, Hickson et al. 1989).
To ensure that the groups fit well within the field-of-view,
we selected those with angular diameters $< \; 7'$.
In redshift, we chose groups with $z \leq 0.05$ so that objects as faint as
$M_R > -16$ (well into the regime of dwarf galaxies) were above our
detection limit.
For a typical redshift, $z = 0.03$ or $cz = 9000 {\rm \; km \;  s^{-1}}$,
our median limiting apparent magnitude of 21.3 corresponds to $M_R = -14.1$
(using ${H}_0 = 75 {\rm \; km \; s^{-1} \; Mpc^{-1}}$, ${q}_0 = 0.1$).
Of the 66 Hickson compact groups that satisfied these criteria,
49 are observable in November from Palomar.

We obtained Johnson R-band images of 39 Hickson compact groups (listed
in Table 1) using the
1.5-m telescope at Palomar.
A thinned Tektronix $2048 \times 2048$ CCD was binned by $2$
in both directions and provided a $12.8' \times 12.8'$
field-of-view with $0.75''/{\rm pixel}$.
Over four nights, two 15-minute exposures were taken of each group.
Offset frames, positioned $10'$ from the group center, were also taken
for 18 HCGs in our sample.
Calibration frames (5-minute exposures of each compact group)
and standard star fields (Landolt 1992) were obtained during the
fifth night under photometric conditions.

The data were reduced using IRAF\footnote{IRAF is distributed by
the National Optical Astronomy Observatories, which are operated
by the Association of Universities for Research in Astronomy, Inc.
(AURA) under cooperative agreement with the National Science
Foundation.}.
A median of 44 bias frames was subtracted from each image.
The median of 55 images, including standard star fields, calibration frames,
and offset frames (taken $10'$ from the group center),
was used to flat field images because it produced a smoother background than
either twilight or dome flats.
Next, we removed bad pixels and cosmic rays, combined
the two 15-minute images of each group, and calibrated the images using the
5-minute photometric exposures and 
standard stars observed at a range of airmasses.

Our standard star reductions indicated that there was a slight color term between
the Landolt (1992) system and our observations.
For dwarf galaxies (which are the primary focus of this study),
a typical color of $(B-R)=1.0$,
yielded a color term error of only 0.01 magnitudes.
There was no color information available for our dwarf candidates,
so we omitted this apparently insignificant correction.
Our standard star calibration uncertainties were $\sim 0.02$ magnitudes.

We used FOCAS (Faint Object Classification and Analysis System) 
(Jarvis \& Tyson 1981) to identify non-stellar objects in the images.
The software generated a catalog of objects on the basis of user-defined
detection parameters.
We required that objects have at least six contiguous pixels
$1.75 \sigma$ above the local background.
These values were set interactively to include
low surface brightness dwarf candidates that were apparent by visual inspection.
FOCAS computed a significance statistic for each object, which
we used as a detection criterion.
By examining detected objects and their significance,
we defined our lowest acceptable significance value to be 1.0.
Those objects classified as galaxies (i.e., extended) were
included in our galaxy counts.


\section{Determination of a Luminosity Function}

A simple statistical technique is used to determine the luminosity function.
For each CCD frame, regions are selected that define the projected area of a
compact group and the surrounding background (Figure 1).
The magnitude distribution of FOCAS-detected galaxies in the background region
allows us to predict the contribution of foreground and background objects to 
galaxy counts observed at each magnitude interval within the compact group area.
By subtracting background counts from total counts, we estimate a magnitude
distribution which is representative
of the compact group galaxies without determining group membership for
individual objects in the field.
Because of the statistical nature of this method,
it is most effective when
data are combined from many images (the sample is presented in
Figures 2a through 2g).
The details of our procedure are described below.

The compact group area on each image is defined in terms of a group radius.
The area of the group is determined by first finding the smallest circle enclosing
the positions of every member with an accordant velocity.
Galaxies which have relative radial velocities within 1000 km/s are said to have
accordant velocities and are presumed to be physically associated (Hickson et
al. 1992).
Some members originally
cataloged by Hickson (1982) have discordant velocities,
implying that they are foreground/background galaxies and we omit them from
the group.
While the group center remains fixed, the circle is expanded to include surface
brightness contours of $R=23\;
{\rm mag/arcsec^2}$ around each member so that the entire galaxy is considered
part of the group.
This procedure defines the group radius, $R_G$.
A background area is defined as that region beyond twice the group radius
extending almost to the edges of the CCD image.
Galaxy counts from the outer regions of 34 CCD images are combined to 
create a ``background'' magnitude distribution.
Because the limiting magnitude is different for each frame, the faintest
magnitude bins are determined by fewer frames and must be corrected so that
the counts in each bin correspond to equal areas.

The resultant background is compared to Tyson's galaxy counts (Tyson 1988)
 to check that we have
extracted an appropriate representation of the field luminosity function
from our data (Figure 3a).
\renewcommand {\thefigure}{\thefiga \alph{figb}}
\setcounter{figa}{3}
\setcounter{figb}{1}
\begin{figure}[t]
 \plotone{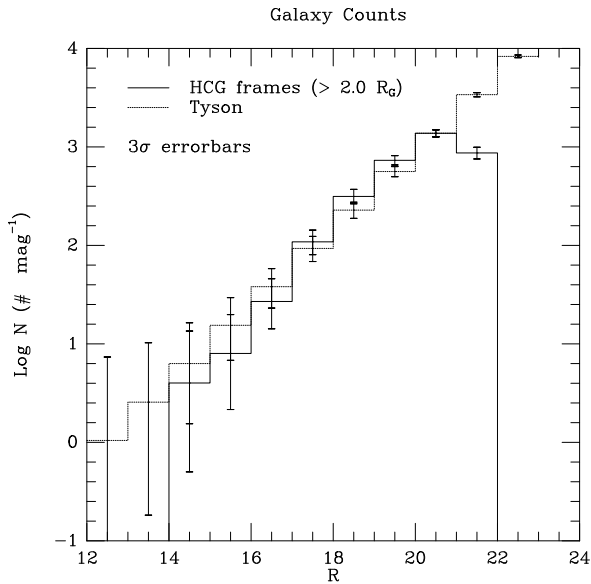}
\caption{{\sc background galaxy counts.} \small
Background galaxy counts using HCG frames are compared to Tyson's counts.
Counts per magnitude interval are plotted as a function of
apparent magnitude ($R$). The expected counts from Tyson's equation
(Tyson 1988) are adjusted to the same area as the background.}
\end{figure}
We find the following deviations of the HCG background from Tyson's counts:
there is an absence of bright galaxies ($R < 14.0$) and an excess in some of the
fainter bins ($R = 18.5, 19.5$).
The lack of bright galaxies is readily explained by one of Hickson's original
selection criteria: between one and three group radii, there can be no galaxies
brighter than the brightest group member.
The result reported by Ramella et al. (1994), that many compact groups are 
embedded in rich looser systems, may be responsible for the excess counts
observed in the fainter bins of our background.
To confirm that these deviations are not a problem with our FOCAS detection
parameters, we compare Tyson's counts to those of our ``offset'' frames.
The offset frames, taken at Palomar during the same run, are single 15-minute
exposures positioned $10'$ from the compact group center.
The background created from 18 offset frames is more similar to Tyson's
counts (Figure 3b) and we conclude that the FOCAS detection procedures are sound,
\setcounter{figa}{3}
\setcounter{figb}{2}
\begin{figure}[t]
 \plotone{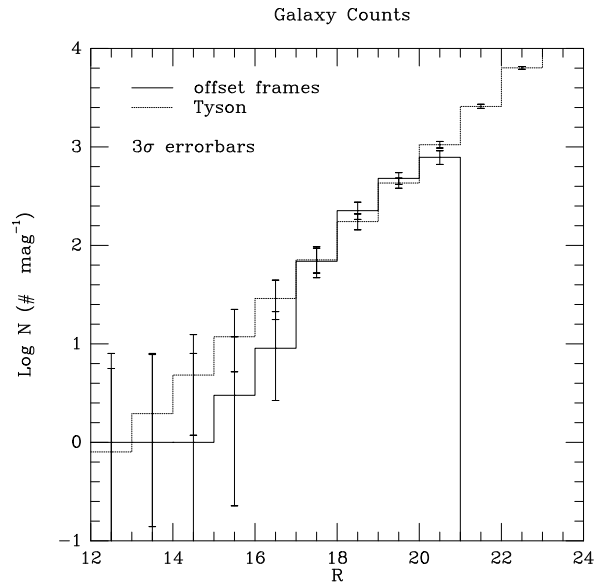}
\caption{{\sc background galaxy counts.} \small
Background galaxy counts using offset frames are compared to Tyson's counts.
Counts per magnitude interval are plotted as a function of
apparent magnitude ($R$). The expected counts from Tyson's equation
(Tyson 1988) are adjusted to the same area as the background.}
\end{figure}
although even the offset frames may still sample the surrounding loose group
because they overlap the HCG frames by $\sim 2.8'$.
If we overestimate the background counts due to this contamination,
we will underestimate the number of dwarf members. 
We could use offset frames that are widely separated from the group
to get better agreement with Tyson's counts, but they might not
properly sample the background at the position of the group.

For each group, the background magnitude distribution is converted to absolute
magnitudes using the redshift of the group.
The area in which the background contributes must also be defined.
We take into account that the background cannot be detected in regions
covered by giant group members.
Also, we visually examine each group and classify some FOCAS-detected objects
as HII regions.
Within eleven groups, we omit a total of 31 small, bright objects located along
spiral arms.
In addition, there are three groups containing four large face-on spirals which
exhibit very clumpy substructure and FOCAS mistakenly identifies this as
several individual objects.
We eliminate such areas from the group analysis. 

We now determine the 
contribution of group members to the galaxy counts within each compact group
area and combine the data from all groups to produce a luminosity function of
HCG galaxies.
The counts in each absolute magnitude interval represent the average number of
galaxies per compact group, although not every group contributes to the
fainter bins because of incompleteness.
To estimate the completeness of each image, detected objects are binned by
apparent magnitude and the faintest bin with counts within $1\sigma (\sqrt{n})$
of the peak of the distribution defines the completeness limit. 
Table 1 lists the limiting apparent magnitude $R$ of each frame and the
corresponding absolute magnitude limit at the redshift of each group.

How well do the
FOCAS magnitudes of the bright members agree with the photometric catalog of
Hickson et al. (1989)?
Of 182 original group members (including those galaxies that were
later discovered to have discordant velocities), 172 galaxies are detected
but 68 of these have saturated pixels in the central region.
Disturbed morphologies and superpositions of stars or other galaxies are the
primary causes of non-detection.
Using the remaining objects, the median difference between the total galaxy
magnitudes calculated from fluxes returned by FOCAS and magnitudes computed
from information in the catalog is $\Delta M_R = 0.06$, with the Hickson
catalog magnitudes usually being brighter.
The average difference is $\Delta M_R = -0.01 \pm 0.37$.
For completeness, we use the catalog magnitudes to determine
the bright end of our luminosity function.

The final critical issue involves a method of normalization.
The ``standard'' procedure begins with a sample completeness correction, i.e.,
determining the magnitude at which a sample becomes incomplete.
Following the example of Hickson et al. (1989), one can calculate this value by
summing the $B_T$ magnitudes of individual galaxies to obtain group magnitudes,
plotting a cumulative distribution of those group magnitudes, and then fitting
the distribution to the equation
$$ N(m)={2n \over  \pi}\arctan[10^{0.6(m-m_0)}] $$
where $n$ is the number of groups, $m$ is a $B_T$ magnitude,
$N(m)$ is the number of groups with magnitude $\le \; m$, and $m_0$
represents the magnitude where the data become incomplete.
This assumes that compact groups are uniformly distributed in space and
one can then express the probability of detecting a group with magnitude $m$ as
$$P(m)=[1 + 10^{1.2(m-m_0)}]^{-1}.$$
Next, the effective volume of each group, i.e., a sphere whose radius is defined
by the maximum distance at which a group is detectable, can be calculated:
$$ V_i = A \int_0^{\infty} P(m_i) r^2 dr $$
where $V_i$ is the volume of the $i^{\rm th}$ group, $A$ is the solid angle of
the survey, $r$ is distance in Mpc, and $P(m_i)$ is defined previously.
A luminosity function is then normalized by dividing galaxy counts for each
group by the effective volume and summing the counts from all groups.
\setcounter{figa}{4}
\setcounter{figb}{0}
\begin{figure}[t]
\plotone{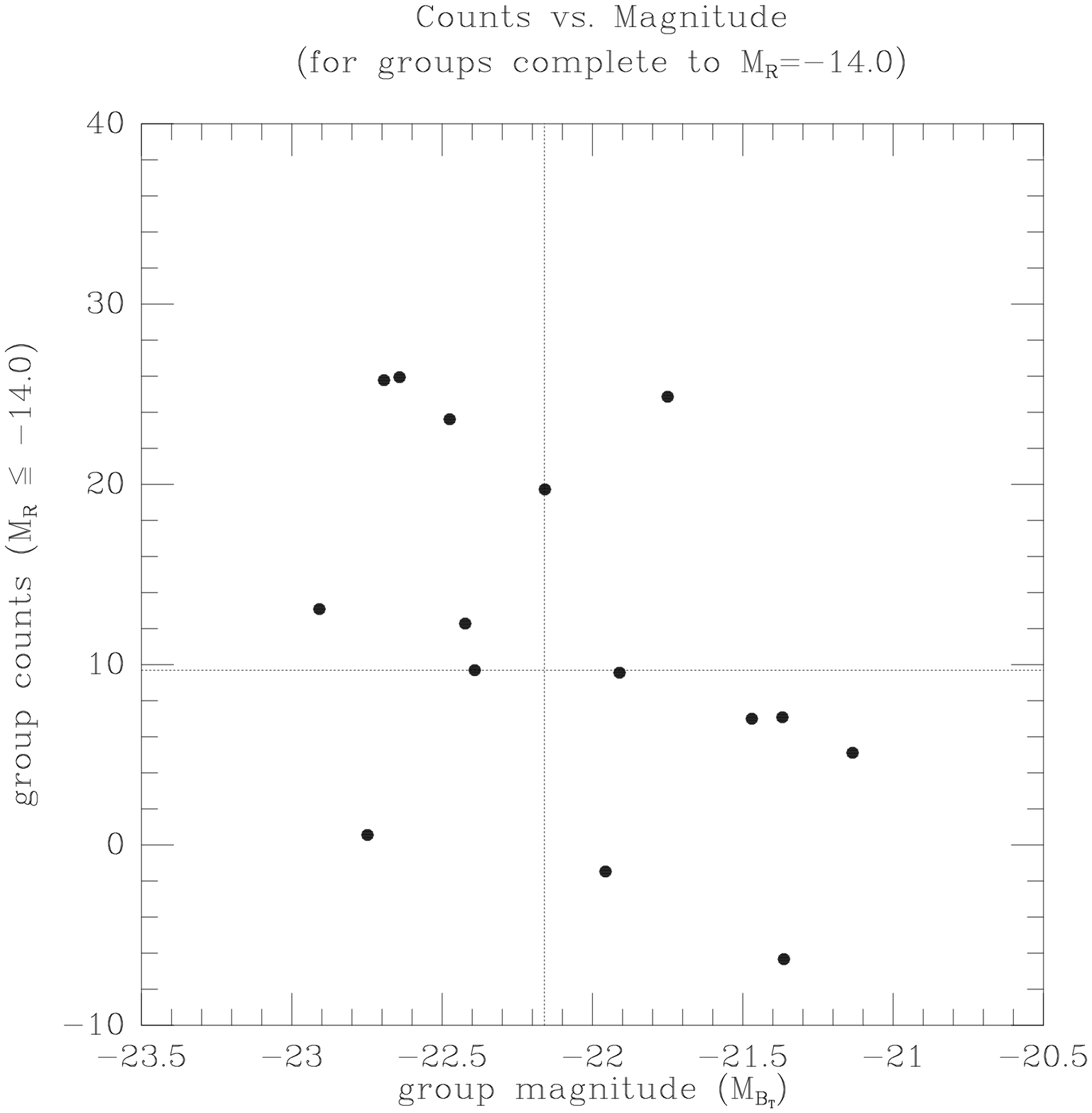}
\caption{{\sc group population vs. group magnitude.} \small
Data points show the number of galaxies brighter than $M_R=-14.0$ for
each group within $R=1.50 R_G$ plotted as a function of group magnitude.
Tidal dwarf candidates are excluded from the counts.
There is a trend for the
brightest groups to have the largest population. The confidence level in this
correlation is 97.0\%. The dotted lines mark the median values.}
\end{figure}

Although this method of normalization is used in determining earlier compact
group luminosity functions (Mendes de Oliveira \& Hickson 1991,
Ribeiro et al. 1994), we normalize our luminosity function by simply
dividing the galaxy counts in each magnitude bin by the number of groups contributing
to that bin.
We justify this new approach by the following argument.
The goal of this study is to develop a luminosity function 
that reflects the membership of a typical compact group as defined by
Hickson, rather than to demonstrate how a hypothetical compact group population
defined more generally contributes to the Universal luminosity function.
Our method weights galaxy counts from each group equally so that the counts
can be averaged.
Galaxies identified in more distant groups should not carry less weight.
Equal weighting enables us to compare directly the luminosity
functions of various groups, for example that of groups with diffuse x-ray emission to that of groups without
emission, when each bin reflects the average number of galaxies expected
in a compact group with (or without) that property.
The luminosity dependent weighting can also distort the luminosity function if
there is a correlation between group luminosity and the dwarf galaxy population.
Using a $2 \times 2$ contingency table, we find a significant correlation (Figure 4)
between group absolute magnitude and number of faint galaxies
(the probability of randomly producing a greater or equal correlation is 3\%).
Such a trend implies that groups having the largest numbers of
dwarf galaxies also have the largest effective volumes, and hence the lowest
weighting factors.
Applying the standard normalization procedure in this situation
systematically reduces the dwarf to giant ratio.
For this reason, we prefer to measure the number of galaxies per group, rather
than per volume.


\section{Results}

We present a luminosity function of compact group galaxies based on ``total''
absolute magnitudes in Figure 5.
\setcounter{figa}{5}
\setcounter{figb}{0}
\begin{figure}[t]
\plotone{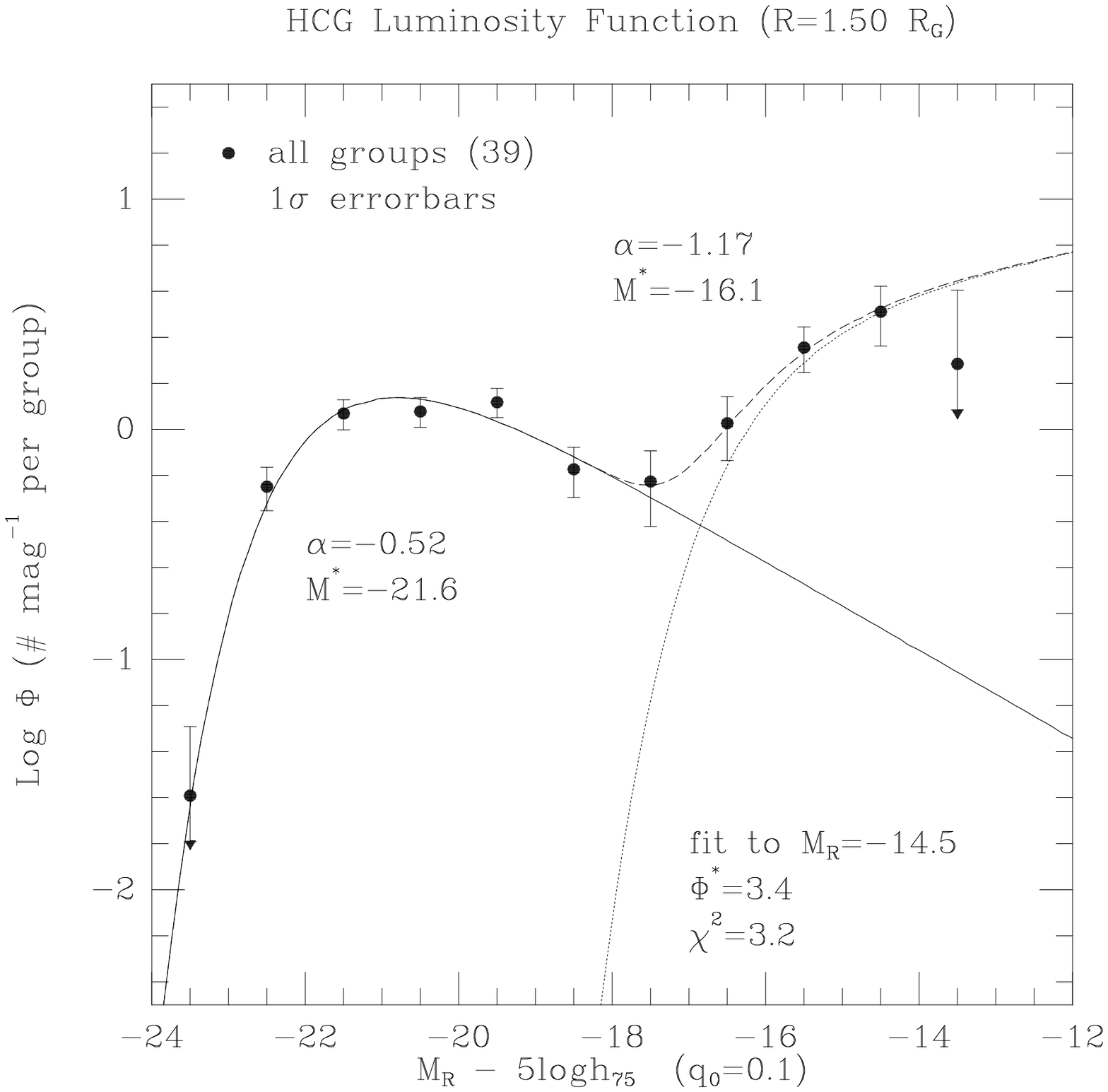}
\caption{{\sc luminosity function of HCG galaxies.} \small
Data points represent the average number of galaxies per group in each magnitude
 bin.
The bright and faint populations are fit separately using 2 Schechter functions.
The solid line is the bright end, the dotted line is the faint end, and the
dashed line is the composite fit.}
\end{figure}
It includes the entire sample of 39 compact groups and the group area is
defined by $R = 1.50 R_G$, where $R_G$ is the group radius.
We adopt this radius because it is the largest area for which all 39 groups
are included in their entirety within our images.
Each point in our luminosity function graphs represents the number of galaxies
per group within a one magnitude interval centered at that point.
The errorbars represent $1 \sigma$ Poisson errors.
As the original counts from group and background regions of different frames
are combined, errors are propagated by the standard method:
${\sigma}_z^2 = a^2{\sigma}_x^2 + b^2{\sigma}_y^2$,
where $a$ and $b$ are numerical constants.

We find that a single Schechter function (Schechter 1976) does
not provide a good fit to the data. 
Using the parameters of the best fit function to define a parent population
($\alpha=-0.72$, $M^*=-21.8$, and ${\Phi}^*=2.6$),
a ${\chi}^2$ test gives values of ${\chi}^2 = 31.2$ with seven degrees of
freedom and $P_{\chi} < .001$
where $P_{\chi}$ is the probability of exceeding ${\chi}^2$ for a given
distribution and its expectation value is 0.50.
The result of the ${\chi}^2$ test indicates very poor agreement between the best
fit function and the data.
We obtain a much better fit (${\chi}^2 = 3.2$ with five degrees of freedom 
and $P_{\chi} \sim 0.67$)
using the composite of two Schechter functions
applied simultaneously to the bright and faint galaxy populations.
The best-fit Schechter functions, shown in Figure 5, have slopes of 
$\alpha = -0.52$ and $\alpha = -1.17$ for the bright and faint ends,
respectively.

It is also useful to examine luminosity functions within several radii (Figure 6).
\setcounter{figa}{6}
\setcounter{figb}{0}
\begin{figure}[t]
\plotone{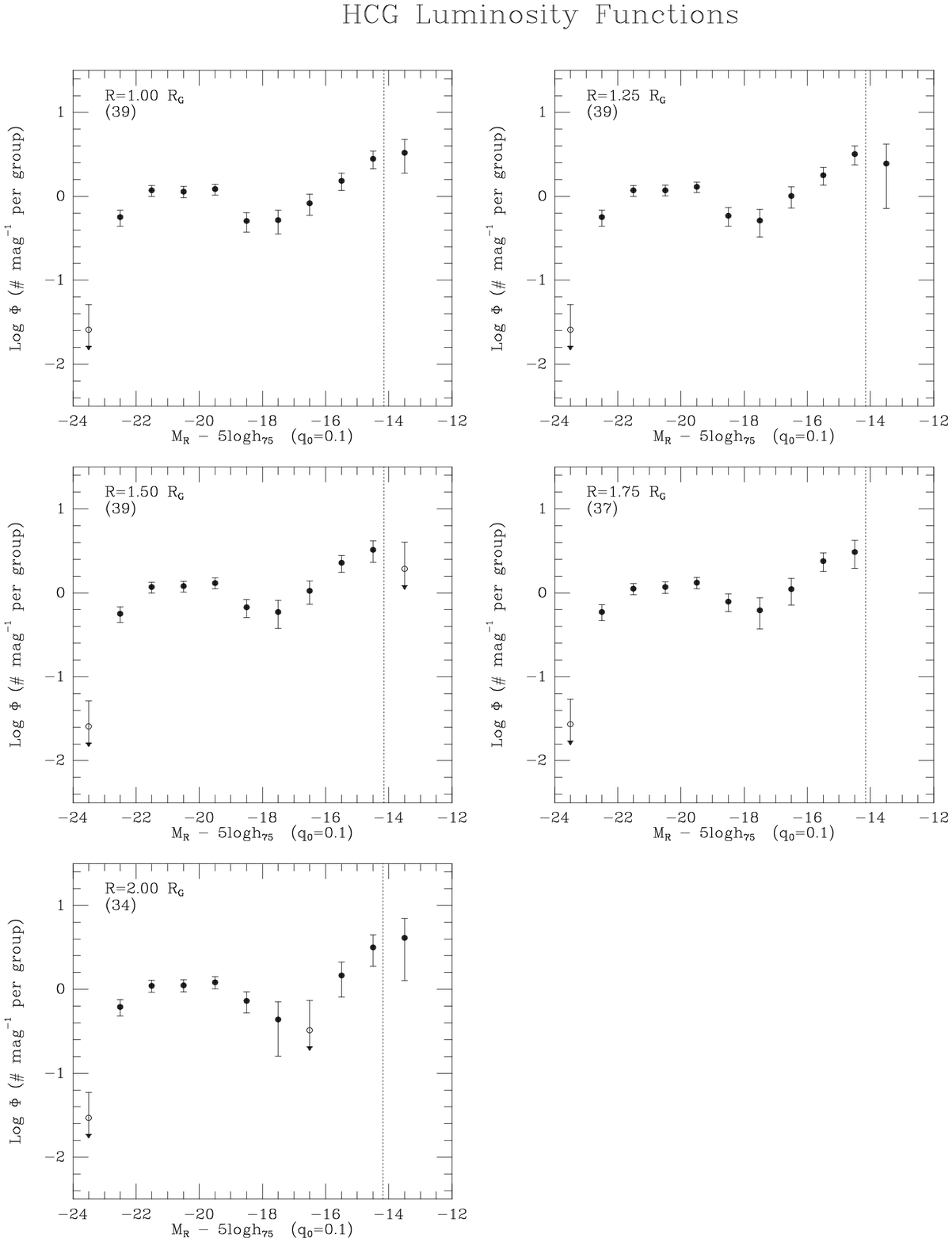}
\caption{{\sc luminosity functions within different group radii.} \small
The number in parentheses in the upper left corner of each plot is the number of
groups in the sample. The vertical lines mark the median completeness for the
groups contributing to the luminosity function.
An open circle denotes the data point is an upper limit.}
\end{figure}
The general shape of the luminosity function is similar for all radii: 
following the data points from bright to faint magnitudes, the
bright end begins to fall off beyond $M_R = -19.5$ and continues until
$M_R = -16.5$ where there is an upturn.
The trough in the luminosity function is unusual, so we examine a few
possible explanations.
The first possibility is that the background has been overestimated at
intermediate magnitudes (recall the excess counts we found relative to
Tyson's counts; Figure 3).
When the apparent magnitudes are converted to absolute magnitudes, the magnitude
bins in which this excess is present are primarily $M_R \ge -17.5$.
Although the trough is lessened when Tyson's counts are used as the
background (Figure 7), it is not completely eliminated and
\setcounter{figa}{7}
\setcounter{figb}{0}
\begin{figure}[t]
\plotone{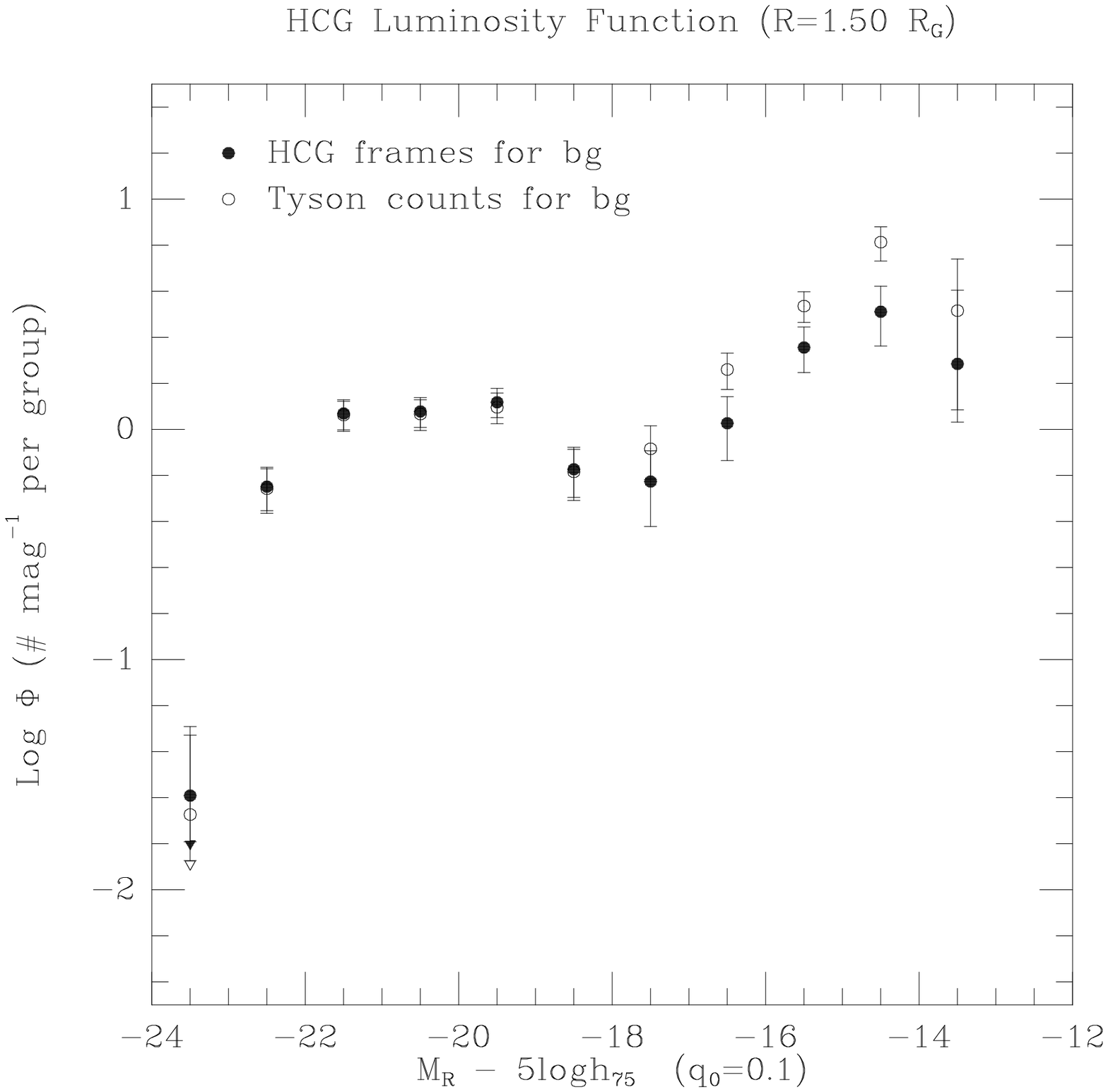}
\caption{{\sc comparison of luminosity functions using Tyson's counts and HCG
frames.} \small
This figure shows the difference in using Tyson's counts and the HCG frames
to calculate the number of background galaxies.
The HCG background has excess counts
(above what is expected from Tyson counts) which leads to fewer compact group
galaxies at the faint end of the luminosity function.
However, the HCG background
may be more representative of regions surrounding compact groups.}
\end{figure}
\setcounter{figa}{8}
\setcounter{figb}{0}
\begin{figure}[t]
\plotone{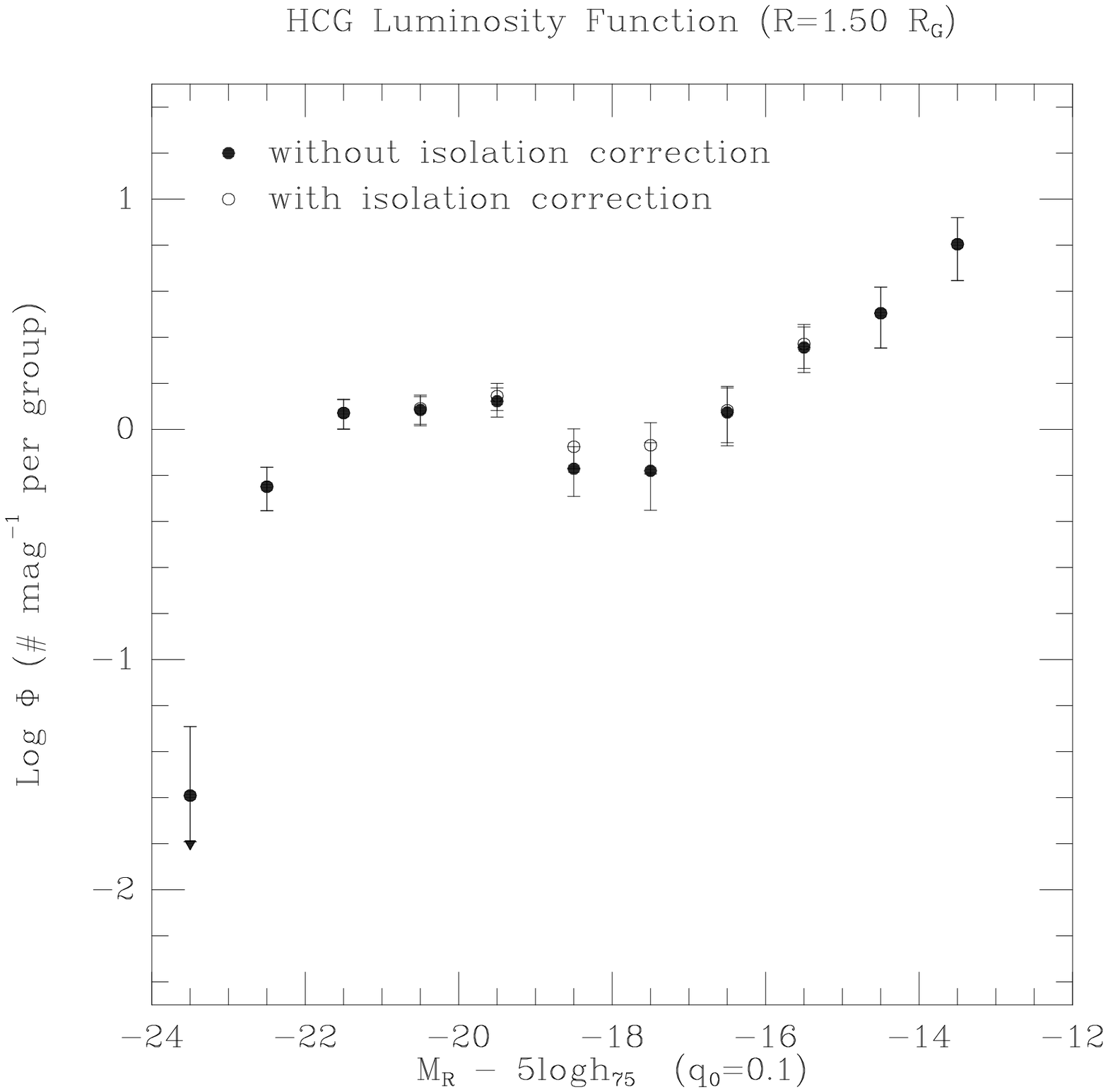}
\caption{{\sc comparison of luminosity functions with and without isolation
correction.} \small
This figure shows the effect of Hickson's original isolation criterion by
accounting for the lack of background galaxies within a certain magnitude range.
Some counts are recovered in the region of the dip in the luminosity function
but isolation is not a major contributor to the deficit of intermediate
luminosity galaxies. The background for these luminosity functions is defined
as the region beyond 3 group radii instead of 2 group radii in other plots.}
\end{figure}
again we emphasize that Tyson's counts are not necessarily representative of
regions surrounding compact groups.
The second possibility is that the original selection criteria imposed by
Hickson (1982) have led to an observational bias.
Recall the isolation criterion: within three group radii, there are no galaxies
(other than group members) within three magnitudes of the brightest group
member.
This means that by definition, the background does not contribute to certain
magnitude bins.
In order to test this possibility, we create a background using counts beyond
three group radii and because there cannot be background galaxies within
three magnitudes of the brightest member, we do not subtract a background
contribution for those magnitude bins.
The corrected luminosity function is shown in Figure 8 and although the trough
is again lessened, it is not removed entirely.
The third possibility we consider is
that there is incompleteness at these magnitudes 
because such galaxies are preferentially located
near the giant members and so are not detected by FOCAS.
Because the deficit at intermediate magnitudes is not seen in certain subsets of the sample,
such as groups with x-ray halos and tidal dwarfs, we conclude that there is a genuine
deficit of intermediate magnitude galaxies in compact groups relative to the luminosity
functions of field and cluster environments.

Because a ``group'' radius probes different physical distances in different groups, it is 
valuable to  examine luminosity functions with radii defined in terms
of kpcs (Figure 9).
\setcounter{figa}{9}
\setcounter{figb}{0}
\begin{figure}[t]
 \plotone{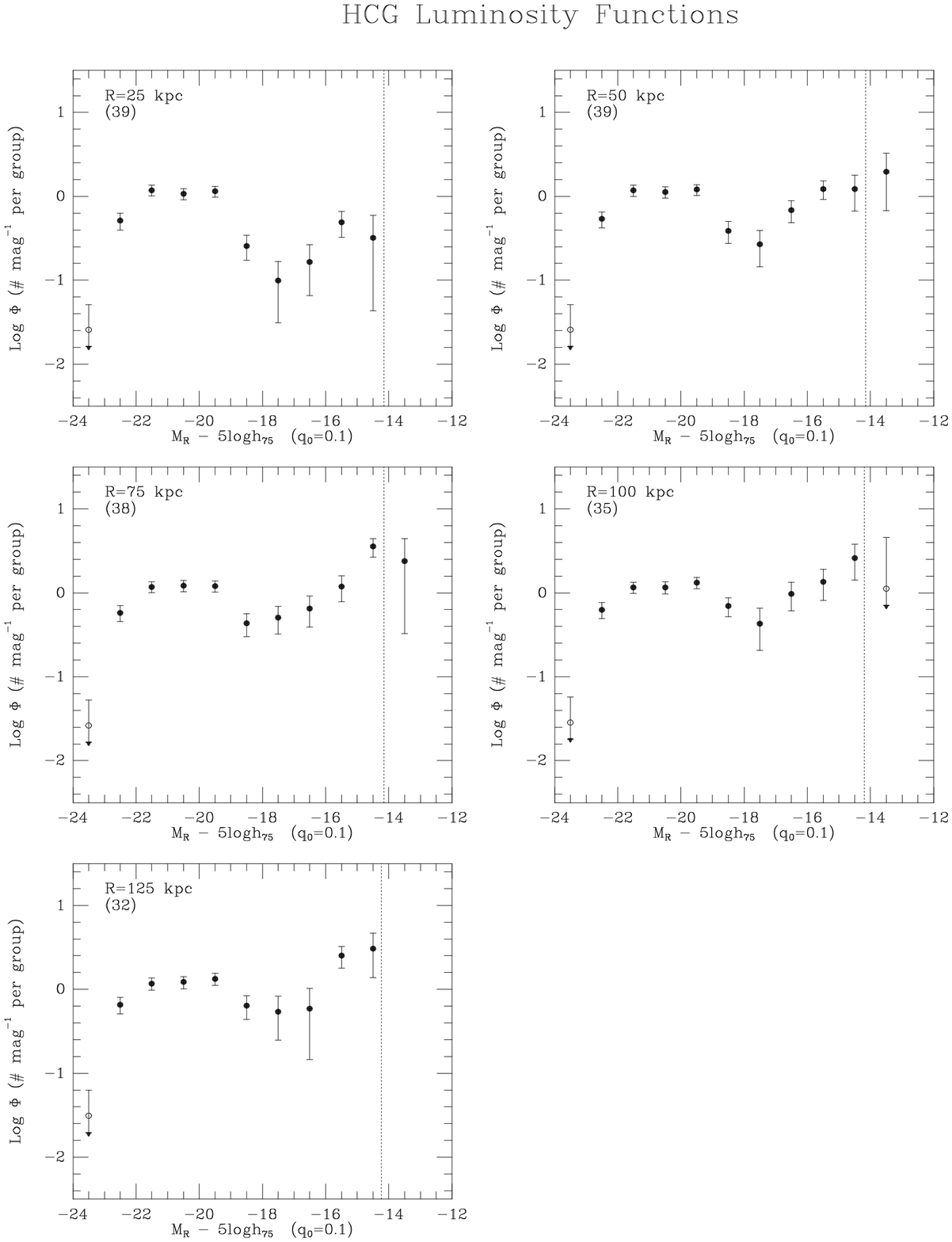}
\caption{{\sc luminosity functions within different physical radii.} \small
The number in parentheses in the upper left corner of each plot is the number of
groups in the sample. The vertical lines mark the median completeness for the
groups contributing to the luminosity function.
An open circle denotes the data point is an upper limit.}
\end{figure}
Again we see the same trends, but the deficit of intermediate galaxies is very evident
at $25{\rm kpc}$ and gradually becomes less pronounced.
This result suggests that the presence of these intermediate galaxies depends
on distance from the group center.
The dwarf galaxy population ($M_R > -18$) is present at all physical radii.

To examine the relationships between dwarf galaxy evolution and environment,
we construct luminosity functions for various group subsets
based on the following properties:
the existence of tidal dwarf candidates (as determined by Hunsberger et al. 1996),
degree of compactness (based on the median separation between giant members;
Hickson 1992),
estimated mass-to-light ratio (from Hickson 1992),
the presence of diffuse x-ray emission (Ponman et al. 1996),
projected velocity dispersion of giant members (Hickson 1992),
fraction of spiral galaxies (Hickson 1989),
type of dominant (first-ranked) galaxy (Hickson 1989),
and number of giant members (Hickson 1992).
These properties are listed for groups in our sample in Table 2:
\newline
\hspace*{2em}{\it Column 1} is the group number with $\ast$ to indicate
tidal dwarf candidates and $\dagger$ to indicate diffuse x-ray emission.
\newline
\hspace*{2em}{\it Column 2} is the number of giant members with accordant
velocities (within 1000 km/s of the group median velocity).
\newline
\hspace*{2em}{\it Column 3} gives the morphological type of the first-ranked galaxy.
\newline
\hspace*{2em}{\it Column 4} is the spiral fraction of giant members. 
\newline
\hspace*{2em}{\it Column 5} lists the estimated mass-to-light ratio (highly
uncertain due to the small number of identified members). 
\newline
\hspace*{2em}{\it Column 6} lists the projected velocity dispersion. 
\newline
\hspace*{2em}{\it Column 7} lists the median separation between giant members. 
\newline
It is impractical to present all these luminosity functions and in many cases,
the presence or absence of a particular property does not produce luminosity
functions that differ significantly from each other.
However, the properties that do substantially impact the
shape of the luminosity function are summarized in Figures 10 $-$ 12 and
discussed below.
\setcounter{figa}{10}
\setcounter{figb}{1}
\begin{figure}[t]
\plotone{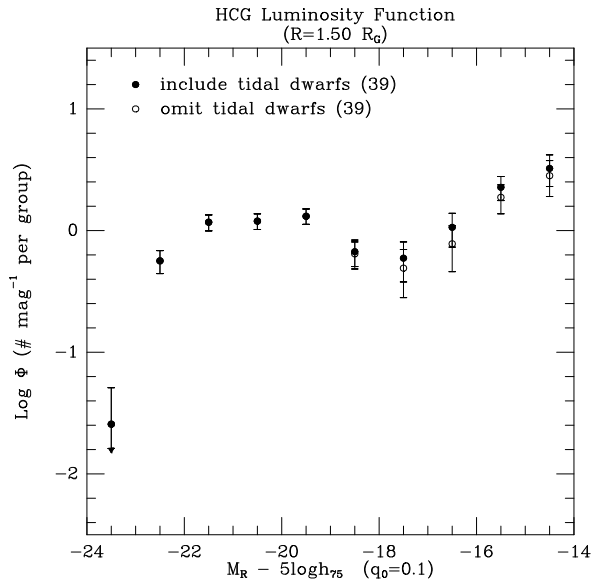}
\caption{{\sc comparison of luminosity functions with and without
tidal dwarf candidates.} \small
This plot shows how tidal dwarfs contribute to galaxy counts averaged over all
groups.
The errorbars for this and subsequent plots are $1\sigma$ and an errorbar
with an arrow indicates the data point is an upper limit.}
\end{figure}
\setcounter{figa}{10}
\setcounter{figb}{2}
\begin{figure}[t]
\plotone{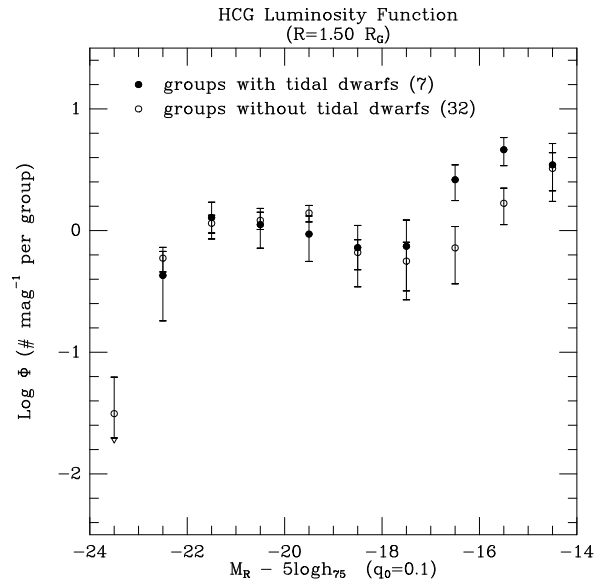}
\caption{{\sc comparison of luminosity functions of groups with and without
tidal dwarf candidates.} \small
One luminosity function is produced by groups which have tidal dwarfs
and the other by those which do not have tidal dwarfs.
The numbers in parentheses at the top
of the plot indicate the number of groups in each subset.}
\end{figure}

First, we examine the relationship between the presence of tidal dwarfs 
and the luminosity function.
Figure 10a shows that the inclusion of tidal dwarf candidates has a negligible
impact on the total luminosity function of galaxies in compact groups.
Figure 10b illustrates that there is a significant difference at the faint end of luminosity
functions for groups with and without tidal dwarfs,
but this difference is due {\it only} to the tidal dwarfs themselves.
To simplify the discussion of how other group properties affect the
luminosity function, we eliminate tidal dwarf candidates from the
galaxy counts in all subsequent luminosity functions.

Second, we examine whether galaxy types of giant group members are correlated 
with the existence of a dwarf population.
If we divide the sample in two, depending on whether the first-ranked galaxy
is elliptical/lenticular or spiral, we find an excess of dwarf galaxies in
the groups with E/S0 first-ranked galaxies (Figure 11a).
\setcounter{figa}{11}
\setcounter{figb}{1}
\begin{figure}[t]
\plotone{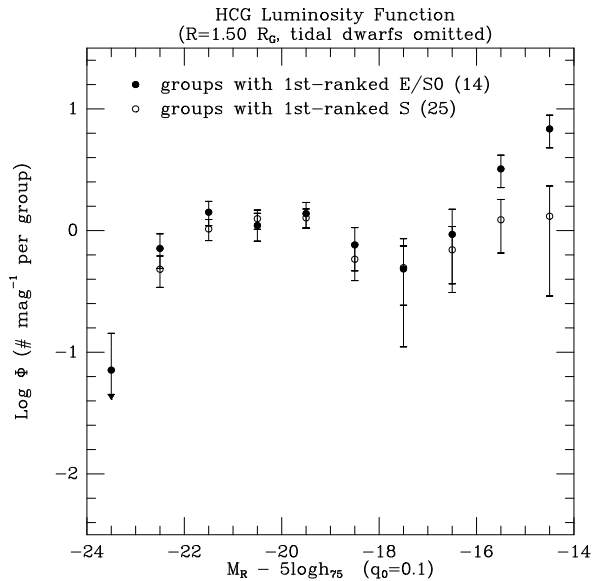}
\caption{{\sc comparison of luminosity functions of groups with and without
a dominant E/S0 galaxy.} \small
One luminosity function is produced by groups having a first-ranked (dominant)
elliptical galaxy and the other by those having a first-ranked spiral.
There is an excess of faint galaxies in the groups with a dominant elliptical.}
\end{figure}
Furthermore, we note that the faint galaxy population still exists when
the compact group area is expanded
to two group radii, $R = 2.00 R_G$ (Figure 11b) 
\setcounter{figa}{11}
\setcounter{figb}{2}
\begin{figure}[t]
\plotone{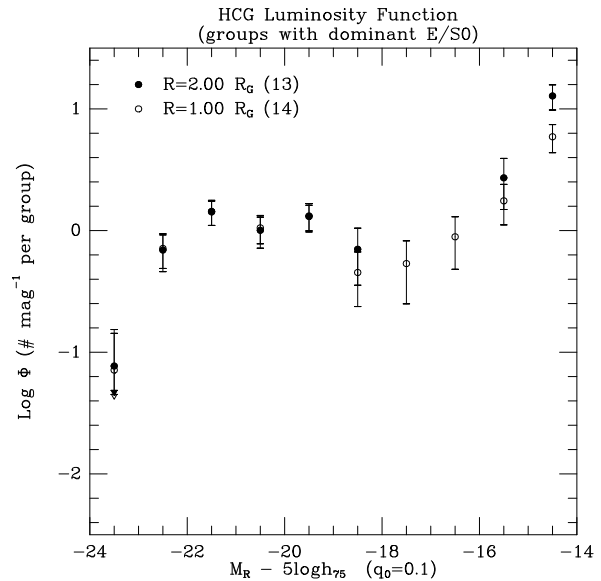}
\caption{{\sc comparison of luminosity functions of dominant E/S0 groups
within different radii.} \small
A dwarf population is still
detected when the group region extends to $R=2.00 R_G$, beyond
the immediate vicinity of the giant members.
The ``missing'' data points in the plot are caused by negative
galaxy counts.
When the region is expanded to a large enough distance, the background
dominates the galaxy counts and negative values for compact group counts
are possible. In this case $\log \Phi$ cannot be plotted.}
\end{figure}
while the intermediate luminosity galaxies are no longer measurable
above the background.
One possible explanation for the apparent excess of 
faint galaxies in the E/SO groups is that these
compact groups inhabit a richer environment.
However, 9 of the 14 E/S0 dominated groups have backgrounds (counts
beyond $R = 2.00 R_G$) for which the galaxies per unit area lie below the average of the
combined frames.
Therefore, the background contribution in the E/S0 groups is, if anything,
actually slightly over-estimated, and so 
we conclude that the faint excess in these groups is associated with the HCGs.

Third, we examine the role of the intergalactic medium.
Because diffuse x-ray emission is associated with a hot intra-cluster medium, we
divide our sample into groups with and without such x-ray emission based on
the results of a ROSAT survey of HCGs (Ponman et al. 1996).
Ponman et al. detected diffuse x-ray emission in 22 of 85 groups.
Seven of these 22 groups are in our sample and the comparison of luminosity
functions of groups with and without detected x-ray halos is presented in Figure 12a.
\setcounter{figa}{12}
\setcounter{figb}{1}
\begin{figure}[t]
\plotone{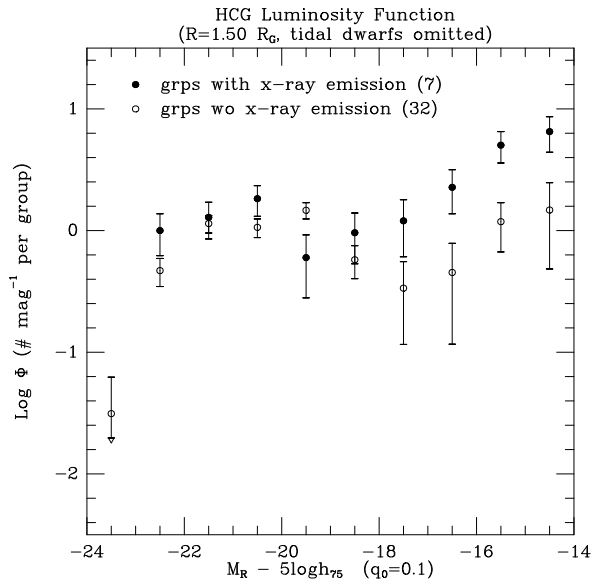}
\caption{{\sc comparison of luminosity functions of groups with and without
diffuse x-ray emission.} \small
There is an excess of dwarfs in the groups with
x-ray halos. Also, the deficit of intermediate galaxies noted in the total
luminosity
function does not appear for groups with x-ray emission.}
\end{figure}
Groups with x-ray emission have a substantially larger dwarf population 
than groups without x-rays.
When we compare the luminosity functions of x-ray groups within different radii
(Figure 12b), we find that the number of faint galaxies increases with radius.
\setcounter{figa}{12}
\setcounter{figb}{2}
\begin{figure}[t]
\plotone{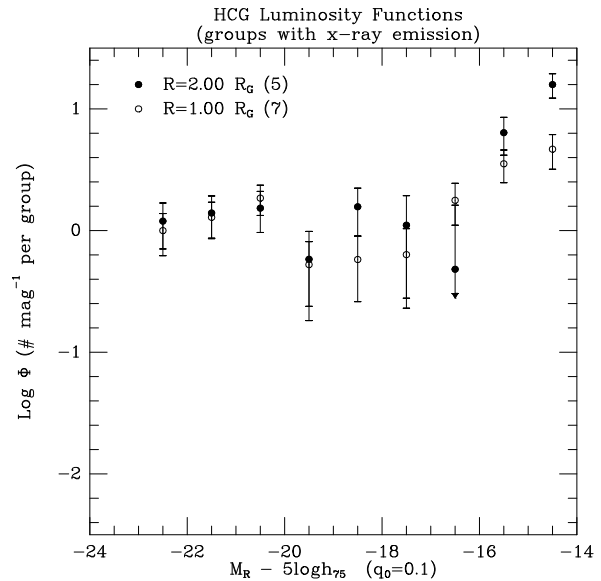}
\caption{{\sc comparison of luminosity functions of x-ray emitting groups
within different radii.} \small
As in the case of groups with dominant ellipticals,
there is a significant dwarf population further out from the giant members.}
\end{figure}

We infer that for x-ray detected groups, the dwarf population extends to a
distance of at least two group radii.
Five of the seven x-ray groups also have first-ranked
E/S0 galaxies and the other two have tidally interacting members.

In Table 3 we summarize the average galaxy counts for several subsets:
\newline
\hspace*{2em}{\it Column 1} describes each subset and the number of groups
is shown in parentheses. 
\newline
\hspace*{2em}{\it Column 2} lists giants per group based on cataloged group
members. 
\newline
\hspace*{2em}{\it Column 3} lists dwarf galaxies per group based on estimated
galaxy counts in the magnitude range $-18.0 < M_R - 5 \log h_{75} < -14.0$.
\newline
\hspace*{2em}{\it Column 4} lists dwarfs per giant member. 
\newline
\hspace*{2em}{\it Column 5} shows the result of a
generalized ${\chi}^2$ test which is used to compare the faint end luminosity
functions of two subsets. 
\newline
$P_{\chi}$ evaluates the probability of two data sets 
being drawn from the same parent population.
The calculated values for $P_{\chi}$ indicate that the following subsets are
significantly different at the $2\sigma$ level:
groups with and without tidal dwarfs, groups with
dominant ellipticals and with dominant spirals, groups with and without
diffuse x-rays, and groups with number of giant members above and below the 
median.
The difference due to number of giant members is likely related to other
properties:
the groups with more than 4 giants tend to have dominant ellipticals or
diffuse x-rays.
If groups with x-ray emission are not included, then the faint end
luminosity functions of
groups with low and with high M/L measurements are also statistically different.
In fact, for groups without diffuse x-rays and with high M/L, a dwarf galaxy
population is not detected.

Finally, we want to compare to other luminosity functions.
Because of different normalizations, comparing faint end slopes is not
necessarily meaningful;
we prefer to examine quantities such as dwarf galaxies per group or dwarfs per
giant galaxy.
There are also problems which arise when comparing luminosity functions with
data taken in different filters and at different limiting magnitudes.
In some sense, $B$ and $R$ filters select objects with different stellar
populations at each magnitude and at the faintest magnitudes, objects detected
in one band might not be detected in another.
One can assume a universal color in order to compare luminosity functions in
different bands, but galaxy colors such as $B-R$ can vary by more than
a magnitude depending on morphological type (Fukugita et al. 1995).
Because we use a statistical method for counting galaxies, we do not identify
individual galaxy members and it becomes impractical to compare our
luminosity  function in $R$ to a luminosity function produced with
$B-$band data.
The other major difficulty is handling different completeness limits.
For example, many field luminosity functions are produced from surveys which
cover a large area of the sky but are not faint enough to probe the regime of
dwarf galaxies.
One option is to extrapolate the Schechter function fit but if there is an
upturn at fainter magnitudes that is not being seen, then the reported faint
end slope does not represent the dwarf galaxy population we wish to compare.
Another option is to perform a comparison based only on galaxy counts within
the ``brighter'' completeness limit.
In this case, we may not be including a significant part of the dwarf galaxy
population and so a term like ``dwarfs per giant'' seems inappropriate.

Considering these limitations, we examine various luminosity functions as
candidates for comparison.
We find no suitable field luminosity functions:
the luminosity functions of the Stromlo-APM Redshift Survey (Loveday et al. 
1992), the CfA Redshift Survey (Marzke et al. 1994), and the Durham/UKST Galaxy
Redshift Survey (Ratcliffe et al. 1998) are based on blue or Zwicky magnitudes
and although the luminosity function for the Las Campanas Redshift Survey 
(Lin et al. 1996) is presented in Gunn$-r$, the Schechter function fit extends
only to $M=-17.5 + 5 \log h$.
We would also like to compare to the previous compact group luminosity function
of Zepf et al. (1997) but it uses $B-$band magnitudes and perhaps more
importantly, they recalculate group radii so that our respective
samples cover very different areas with respect to the giant member galaxies.
It is easier to find candidates for comparison among the many
cluster luminosity functions.
Secker et al. (1997) present a luminosity function in $R$ for the core of
the Coma Cluster which extends well beyond the completeness limit needed for
a suitable comparison to our luminosity function of compact group galaxies.
Based on the luminosity function shown in Figure 9 of the Secker et al. (1997)
paper, we calculate a dwarf to giant galaxy ratio  $\sim 4.2 \pm 0.5$ for Coma.
This value is more than double the number of dwarfs per giant in the
``typical'' compact group but it is comparable to the dwarfs per giant in
compact groups with diffuse x-ray emission.
We further note that the general shape of the Coma luminosity
function (Secker et al. 1997, Trentham 1998, Jorgensen \& Hill 1997)
is similar to that of HCG galaxies. 
There is a slight dip at intermediate luminosities and then a sharp
upturn creating a faint end with a steep slope.
Such behavior is also seen in the rich cluster Abell 665 (Wilson et al. 1997).


\section{Summary \& Discussion}

Summarizing, we find that

\begin{itemize}
\item Dwarf galaxies are detected in Hickson compact groups using a statistical
method for background subtraction.
The resultant luminosity function compiled from all groups is not
described by a single Schechter function.
When we fit the distribution with a composite of two Schechter functions
(for bright and faint galaxies), we find that the faint end slope
is $\alpha = -1.17$, down to a limiting magnitude of
$M_R=-14.0 + \log h_{75}$.
\item The general shape of the luminosity function exhibits a dip at
$M_R=-17.5 + \log h_{75}$ and $\alpha = -0.52$ for the Schechter function
characterizing the bright galaxy population.
The decreasing number of galaxies per magnitude with decreasing luminosity
suggests a ``deficit'' of intermediate luminosity galaxies in comparison
to most field luminosity functions ($\alpha \sim 1.0$).
\item Dwarf galaxies presently observed in tidal debris (tidal dwarf candidates) make
a small contribution to the total luminosity function of all groups, but groups with tidal
dwarfs have a significantly larger dwarf population than
those without tidal dwarfs.
\item Groups with first-ranked E/S0 galaxies have an excess faint galaxy
population over those with first-ranked spirals. 
\item Groups with diffuse x-ray emission have a large dwarf population
extending to two group radii.
Of the 14 groups with dominant E/S0 galaxies, five of them also have detected x-ray
halos.
\item The dwarf-to-giant ratio in compact groups with x-ray halos is comparable
to that of the cores of rich clusters.
\end{itemize}

To understand these results, we hypothesize the following (realizing that
these are not unique interpretations of the data):
\parindent=0em

1) {\it X-ray halos are an indication of previous or current giant galaxy
interactions.}
\newline
A hot intra-group medium is characterized by extended, diffuse x-ray emission
rather than discrete sources which are associated with individual galaxies.
Both Ebeling at al. (1994) and Pildis et al. (1995) detected extended x-ray
emission in HCGs and discovered a relationship between x-ray emission and the
morphological type of the dominant galaxy, i.e., an x-ray  group usually has
a first-ranked E or S0 galaxy. 
They also reported an anti-correlation between the spiral fraction 
of the group and the presence of an x-ray halo.
A more complete survey of HCGs by Ponman et al. (1996) finds x-ray 
emission in some spiral-rich groups as well.
Recall that five of the seven x-ray groups in our sample have
first-ranked E/S0 galaxies and a low spiral fraction ($< 50\%$) and the other
two are spiral-rich and currently interacting as evidenced by tidal tails.
If one assumes that elliptical and lenticular galaxies are merger remnants, then
this naturally suggests a connection between interactions/mergers and
x-ray halos.
During the merger process, gas and stars from the outer spiral disks are being
tidally stripped.
Supernovae from tidally-triggered star formation and shock-heating in tidal
tails may be responsible for heating the gas to x-ray temperatures.
Models of merging galaxies (Mihos et al. 1998) indicate that low-mass or
extended galaxy halos allow tidal debris to be expelled to large distances.

2) {\it Many dwarf galaxies form in tidal debris during interactions.}
\newline
There is observational and theoretical evidence to support the idea of
dwarf galaxy formation in the tidal debris of giant galaxy interactions
(cf \S1).
In a previous paper (Hunsberger et al. 1996), we compiled a list of tidal
dwarf candidates for our sample of HCGs using FOCAS detections coincident
with tidal features.
We identifed 47 possible tidal dwarfs in seven of the groups.
These tidal dwarfs are responsible for the excess of faint galaxies in
Figure 11b, but the real question is what is the ultimate fate of these
objects?
Elmegreen et al. (1993) predict that if the perturbing galaxy has a greater
mass, the tidal dwarfs will be ejected into the group at large or
become satellites of the new galaxy instead of falling back into the merger
remnant.
Therefore, we expect a large fraction of these systems to survive.

Next, consider the similarity of the luminosity functions of groups
with tidal dwarfs and groups with diffuse x-ray emission.
Neither luminosity function exhibits the deficit of intermediate luminosity
galaxies, and both have significant dwarf galaxy populations.
If x-ray halos are a sign of recent interaction and if tidal dwarfs are
expelled from giants and can
survive for as long as the cooling timescale of the gas then
tidal dwarfs provide a reasonable explanation for the abundant dwarf
population observed in x-ray groups.

3) {\it The size and distribution of dark matter halos profoundly
affects the evolution of groups of galaxies, especially the dwarf population.}
\newline
If individual galaxies have massive halos then mergers occur quite quickly
because of dynamical friction (Barnes 1985).
If the galaxies have low-mass halos then they present smaller cross-sections
for interaction so that the merger rate within the group is slower.
In general, as the mass fraction of a common group halo increases,
the merger rate decreases (Bode et al. 1993).
The mass distribution of a group halo changes as the group evolves.
As mergers occur in the central region, the galaxies' orbital energy is
transferred to the halo causing it to become less centrally concentrated.
Is the dark matter in compact groups in individual galaxy halos or
a common group halo?
The high spatial density of galaxies in
compact groups makes it unlikely that a member can retain an extended halo.
The $M/L$ measurements for each HCG listed in Table 2
are based on velocity dispersions of a few giant galaxy members in each
group which introduces considerable inaccuracy.
Such estimates are meaningful only within the group radius and high M/L values
can mistakenly be inferred from large velocity differences of galaxies
viewed in projection.
Furthermore, it is quite possible that compact groups are not virialized
systems.
Although the reported mass-to-light ratios may be unreliable, we proceed with
a hypothetical discussion.
If a group has a low $M/L$, i.e., below the median,
then galaxies have low-mass halos and either
a) there is a low-mass group halo or
b) the common halo is more massive but extended and not centrally concentrated.
In this situation, giant galaxy interactions can produce tidal
tails (Mihos et al. 1998) and tidal debris (and tidal dwarfs)
can be ejected into an intra-group medium.
In the case of b), the expelled gas should trace the gravitational potential
which is now
shallow but more extended and be detectable as a diffuse x-ray halo.
When a group has a high $M/L$ it means that 
a) the galaxies have massive halos, b) there is a massive, extended group halo,
or c) the common halo is less massive but has a strong central concentration.
For a) or c),
the massive halo(s) prevent formation of tidal tails during an
interaction (Mihos et al. 1998).
So as dwarfs are either cannibalized by massive galaxy halos
or destroyed during a merger of giant members, there is no way
to replenish the population.

4) {\it There exists some mechanism which preferentially eliminates intermediate
luminosity galaxies.}
\newline
First we examine what is known about compact group formation.
Previous studies hint at a connection between compact groups and loose groups.
Despite the original isolation criterion for HCG selection, Ramella et al.
(1994) report that 29 of 38 HCGs are embedded in rich, looser systems.
N-body simulations (Diaferio et al. 1994) suggest that compact configurations
resembling HCGs are continually forming during the collapse of rich loose
groups.
Furthermore, a plot of group magnitude vs. group diameter (Sulentic \&
Raba\c{c}a 1994) for both loose groups from the CfA survey (Geller \& Huchra
1983) and HCGs shows a very smooth transition in parameter space from one sample
to the other.
If a compact group evolves from a looser system, then its initial population
should be similar to giant field galaxies and their companions.

Statistical analyses of the distribution of satellites around giant galaxies
(Lorrimer et al. 1994, Loveday 1997) indicate that faint companions are more
strongly clustered about the primary galaxy than their brighter counterparts.
It is not clear whether a variation of clustering strength or of galaxy counts
is responsible for the lack of intermediate luminosity objects in the luminosity
function of compact group galaxies, but in either case, ``dynamical friction''
offers a plausible explanation. 

In the field and loose groups, giant galaxies have companions orbiting within
large dark matter halos.
Dynamical friction provides a means to decrease the orbital angular momentum of
satellites and the timescale for energy loss is inversely proportional to 
satellite mass.
Because the orbits of more massive companions decay more rapidly, they merge
sooner with the primary galaxy.
If luminosity is proportional to mass, then such a scenario is consistent
with fewer bright companions than faint companions, especially near a giant.
Recall, however, that the dip in the luminosity function is less severe for
groups with tidal dwarfs and essentially disappears for groups with x-ray halos.
(As an aside here, we note that 18\% of the groups in our total sample are
groups with diffuse x-rays while 36\% of the Ribeiro et al. (1994) sample
are x-ray groups.
This may explain why the trough is not a strong feature in their luminosity
function.)
Is it possible that tidal dwarfs replenish the population of intermediate
luminosity galaxies ($-19 < M_R < -17$) and/or that formation of
an x-ray halo inhibits cannibalism?
The tidal dwarf formation models of Barnes \& Hernquist (1992) and Elmegreen
et al. (1993) predict masses in the range $10^7$ to $10^9M_{\odot}$.
For $M/L = 1$, this corresponds to $-18.2 \leq M_R \leq -13.2$ which agrees
with the magnitude range of tidal dwarf candidates in our sample (see Fig. 10a).
The tidal dwarf galaxies cataloged by Duc \& Mirabel (1998) exhibit a magnitude
range of $-18.8 \leq M_B \leq -12.1$ with a median value of $M_B = -14.2$
implying that most tidal dwarfs are low luminosity
objects although a few could be considered intermediate luminosity galaxies.
Now assume that diffuse x-rays trace a common dark
matter halo which is less concentrated than individual galaxy halos
(as discussed previously).
Because the dynamical friction timescale is inversely proportional to
medium density, the loss of orbital energy proceeds more slowly.
Once a significant group potential forms, merging times increase as well as the
expected lifetimes of dwarf members.

5) {\it Groups with first-ranked E/S0 galaxies are more evolved than
other compact groups, i.e., they have undergone previous mergers.}
\newline
In groups with a first-ranked E/S0 galaxy,
there is a substantial population of dwarf galaxies which extends
out to two group radii (Figure 12b).
A possible explanation is that
the dwarfs exist ``at large'' in the group and an 
example of such a free-floating dwarf population is reported in the 
Virgo cluster (Ferguson 1992).
A merger scenario can explain how the dwarf population gets redistributed
and enhanced.
Assume that a dominant-spiral compact group is evolving
into a dominant-E/S0 group. 
Initially dwarfs are bound to individual galaxies and their 
distribution traces the giant galaxy halos.
As two giant members merge, some of the dwarf companions are destroyed,
but more importantly the dark matter is restructured.
It becomes less centrally concentrated and more extended, either creating a
common group halo or making the existing one more massive.
If tidal tails form during the interaction and if they are sites of dwarf
galaxy formation (tidal dwarfs), then there is a mechanism by which the
dwarf population can be replenished and enhanced.
The conditions which allow tidally stripped gas to form a hot (x-ray emitting)
intra-group medium may also distribute tidal dwarfs throughout the group.
When such a merger is complete, the distribution of both the hot x-ray gas
and the dwarf population will trace the extended group halo.
This scenario is consistent with results from a survey of twelve poor groups,
including three HCGs, conducted by Zabludoff \& Mulchaey (Zabludoff \& Mulchaey
1997, Mulchaey \& Zabludoff 1997).
They discovered that nine x-ray detected groups had $20-50$ dwarf members 
($-16 < M_B - 5\log h < -14$) and established that groups with a central
elliptical have an x-ray halo extending $100-300 h^{-1}{\rm kpc}$.
They concluded that the longevity of the dwarf members is increased
because of the distribution of dark matter in an extended common halo.

\parindent=2em
\bigskip

Finally, combining these hypotheses, we speculate that the evolution of a
typical HCG may proceed as follows:
The group forms because dynamical friction of massive dark halos around field
galaxies brings them together.
These newly formed HCGs contain dwarf companions to the giant galaxies, but
these are gradually cannibalized as interactions occur.
Eventually a major merger occurs and this has several consequences:
the formation of a giant E/S0 galaxy, the production of an x-ray halo, and
the formation of tidal dwarfs.
The ongoing interactions strip gas, dark matter, and dwarf companions from the
giants and spread them into a common group halo.
This redistribution extends the lifetime of the dwarf population 
(and the group as a whole) and creates
an environment more conducive to the tidal dwarf replenishment process.

\acknowledgments
We would like to thank M. Bershady, C. Churchill, R. Ciardullo, J. Feldmeier,
J. Hibbard, and P. Hickson
for their helpful comments.
SDH and JCC acknowledge financial support from NSF grant AST-9529242.
DZ acknowledges financial support from an NSF grant (AST-9619576), a NASA
LTSA grant (NAG-5-3501), the David and Lucile Packard Foundation, and the
Alfred P. Sloan Foundation.

\clearpage
\pagestyle{empty}
\textheight=10.0in
\topmargin=-1.1in
\begin{deluxetable}{cccc}
\tablewidth{3.0in}
\tablenum{1}
\tablecaption{Completeness Limits}
\tablehead{
\colhead{HCG} &
\colhead{z} &
\colhead{$R$} &
\colhead{$M_R$}}
\startdata
001 & 0.0339 &  21.50 & $-$14.19 \nl
003 & 0.0255 &  21.10 & $-$13.97 \nl
004 & 0.0280 &  21.00 & $-$14.27 \nl
005 & 0.0410 &  21.50 & $-$14.61 \nl
006 & 0.0379 &  20.70 & $-$15.24 \nl
007 & 0.0141 &  21.70 & $-$12.07 \nl
012 & 0.0485 &  20.80 & $-$15.68 \nl
013 & 0.0411 &  21.50 & $-$14.62 \nl
014 & 0.0183 &  21.20 & $-$13.14 \nl
016 & 0.0132 &  21.20 & $-$12.42 \nl
020 & 0.0484 &  21.60 & $-$14.88 \nl
024 & 0.0305 &  21.30 & $-$14.16 \nl
025 & 0.0212 &  21.70 & $-$12.96 \nl
026 & 0.0316 &  21.30 & $-$14.24 \nl
028 & 0.0380 &  21.50 & $-$14.44 \nl
030 & 0.0154 &  21.60 & $-$12.36 \nl
031 & 0.0137 &  21.50 & $-$12.21 \nl
032 & 0.0408 &  21.20 & $-$14.90 \nl
033 & 0.0260 &  21.10 & $-$14.01 \nl
034 & 0.0307 &  21.40 & $-$14.07 \nl
037 & 0.0223 &  21.60 & $-$13.17 \nl
038 & 0.0292 &  20.60 & $-$14.76 \nl
040 & 0.0223 &  21.30 & $-$13.47 \nl
043 & 0.0330 &  21.40 & $-$14.23 \nl
046 & 0.0270 &  21.10 & $-$14.09 \nl
047 & 0.0317 &  20.50 & $-$15.04 \nl
049 & 0.0332 &  21.30 & $-$14.35 \nl
051 & 0.0258 &  21.70 & $-$13.39 \nl
052 & 0.0430 &  20.50 & $-$15.72 \nl
056 & 0.0270 &  21.30 & $-$13.89 \nl
089 & 0.0297 &  21.00 & $-$14.40 \nl
092 & 0.0215 &  20.90 & $-$13.79 \nl
094 & 0.0417 &  21.40 & $-$14.75 \nl
095 & 0.0396 &  20.80 & $-$15.24 \nl
096 & 0.0292 &  21.60 & $-$13.76 \nl
097 & 0.0218 &  20.80 & $-$13.92 \nl
098 & 0.0266 &  21.00 & $-$14.16 \nl
099 & 0.0290 &  21.20 & $-$14.15 \nl
100 & 0.0178 &  21.60 & $-$12.68 \nl
\enddata
\end{deluxetable}
\clearpage
\textheight=8.0in
\topmargin=0.0in
\begin{deluxetable}{lcccrrr}
\tablewidth{4.5in}
\tablenum{2}
\tablecaption{Properties of Observed Compact Groups}
\tablehead{
\colhead{HCG} &
\colhead{ } &
\colhead{First} &
\colhead{Spiral} &
\colhead{M/L} &
\colhead{Vel} &
\colhead{Median} \nl
\colhead{\#} &
\colhead{N} &
\colhead{Ranked} &
\colhead{Frac} &
\colhead{$h^{-1}$} &
\colhead{Disp} &
\colhead{Sep} \nl
\colhead{ } &
\colhead{ } &
\colhead{ } &
\colhead{ } &
\colhead{ } &
\colhead{(km/s)} &
\colhead{($h^{-1}$kpc)}}
\startdata
001$\ast$        & 4 & S  & 1/4 &  15 &  85 & 49.0 \nl
003              & 3 & S  & 2/3 & 363 & 251 & 77.0 \nl
004              & 3 & S  & 1/3 & 229 & 339 & 57.0 \nl
005              & 3 & S  & 1/3 &  20 & 148 & 25.7 \nl
006              & 4 & S0 & 2/4 &  60 & 251 & 25.1 \nl
007              & 4 & S  & 4/4 &  14 &  89 & 45.6 \nl
012$\dagger$     & 5 & S0 & 2/5 &  74 & 240 & 58.9 \nl
013              & 5 & S  & 1/5 &  39 & 182 & 46.8 \nl
014              & 3 & S  & 2/3 &  27 & 331 & 26.9 \nl
016$\ast\dagger$ & 4 & S  & 2/4 &  22 & 123 & 44.6 \nl
020              & 5 & S0 & 0/5 &  78 & 275 & 31.4 \nl
024              & 5 & S0 & 2/5 &  40 & 200 & 29.5 \nl
025              & 4 & S  & 2/4 &   9 &  62 & 47.9 \nl
026$\ast$        & 7 & S  & 1/7 &  71 & 200 & 31.6 \nl
028              & 3 & S  & 1/3 &   7 &  85 & 21.9 \nl
030              & 4 & S  & 3/4 &  11 &  72 & 51.3 \nl
031$\ast$        & 3 & S  & 2/3 &   1 &  66 &  8.1 \nl
032              & 4 & E  & 1/4 &  41 & 209 & 61.7 \nl
033$\dagger$     & 4 & E  & 1/4 &  46 & 155 & 24.5 \nl
034              & 4 & E  & 2/4 & 100 & 316 & 15.5 \nl
037$\dagger$     & 5 & E  & 2/5 & 123 & 398 & 28.8 \nl
038$\ast$        & 3 & S  & 2/3 & $-$ &  13 & 58.9 \nl
040              & 5 & E  & 3/5 &  15 & 148 & 15.1 \nl
043              & 5 & S  & 4/5 & 155 & 224 & 58.9 \nl
046              & 4 & E  & 1/5 & 479 & 324 & 39.8 \nl
047              & 4 & S  & 3/4 &   1 &  43 & 36.3 \nl
049              & 4 & S  & 2/4 & $-$ &  34 & 12.3 \nl
051$\dagger$     & 5 & E  & 2/5 &  72 & 240 & 58.9 \nl
052              & 3 & S  & 3/3 & 110 & 182 & 87.1 \nl
056              & 5 & S  & 2/5 &  26 & 170 & 21.4 \nl
089              & 4 & S  & 4/4 &   7 &  55 & 58.9 \nl
092$\ast\dagger$ & 4 & S  & 4/4 &  44 & 389 & 28.2 \nl
094              & 7 & E  & 1/7 & 159 & 479 & 57.5 \nl
095              & 4 & E  & 3/4 &  50 & 309 & 30.2 \nl
096$\ast$        & 4 & S  & 2/4 &  15 & 132 & 30.2 \nl
097$\dagger$     & 5 & E  & 2/5 & 348 & 372 & 63.1 \nl
098              & 3 & S  & 1/3 &  23 & 120 & 27.5 \nl
099              & 5 & S  & 2/5 &  50 & 263 & 42.7 \nl
100              & 3 & S  & 3/3 &  32 &  89 & 38.0 \nl
\enddata
\tablenotetext{} {$\ast$ indicates a group containing tidal dwarf candidates
(Hunsberger et al. 1996)}
\tablenotetext{} {$\dagger$ indicates a group with diffuse x-ray emission
(Ponman et al. 1996)}
\end{deluxetable}
\clearpage
\begin{deluxetable}{lcrrr}
\tablewidth{6.5in}
\tablenum{3}
\tablecaption{Compact Group Populations}
\tablehead{
\colhead{ } &
\colhead{Giants} &
\colhead{Dwarfs$^\ast$} &
\colhead{Dwarfs} &
\colhead{ } \nl
\colhead{Subset} &
\colhead{per} &
\colhead{per} &
\colhead{per} &
\colhead{ } \nl
\colhead{Description} &
\colhead{Group} &
\colhead{Group} &
\colhead{Giant} &
\colhead{$P_{\chi}$$^\dagger$}}
\startdata
all groups$^\ddagger$ &&&&\nl
\hspace*{2em}$R=1.00 R_G$ (39) & 4.2 & $5.7 \pm 0.8$ & $1.4 \pm 0.2$ \nl
\hspace*{2em}$R=1.25 R_G$ (39) & 4.2 & $6.5 \pm 1.0$ & $1.5 \pm 0.2$ \nl
\hspace*{2em}$R=1.50 R_G$ (39) & 4.2 & $7.2 \pm 1.1$ & $1.7 \pm 0.3$ \nl
\hspace*{2em}$R=1.75 R_G$ (37) & 4.2 & $7.2 \pm 1.4$ & $1.7 \pm 0.3$ \nl
\hspace*{2em}$R=2.00 R_G$ (34) & 4.1 & $5.4 \pm 1.5$ & $1.3 \pm 0.4$ \nl
 & & & & \nl
\hspace*{2em}R=25kpc (39)      & 4.2 & $ 1.1 \pm 0.3$ & $ 0.3 \pm 0.1$ \nl
\hspace*{2em}R=50kpc (39)      & 4.2 & $ 3.4 \pm 0.7$ & $ 0.8 \pm 0.2$ \nl
\hspace*{2em}R=75kpc (38)      & 4.2 & $ 5.9 \pm 1.0$ & $ 1.4 \pm 0.2$ \nl
\hspace*{2em}R=100kpc (35)     & 4.3 & $ 5.4 \pm 1.4$ & $ 1.3 \pm 0.3$ \nl
\hspace*{2em}R=125kpc (32)     & 4.3 & $ 6.7 \pm 1.9$ & $ 1.6 \pm 0.4$ \nl
\hspace*{2em}R=150kpc (27)     & 4.2 & $-1.6 \pm 2.9$ & $-0.4 \pm 0.7$ \nl
 & & & & \nl
 & & & & \nl
with tidal dwarfs$^\ddagger$ (7) & 4.1 & $11.4 \pm 2.3$ & $2.8 \pm 0.6$
 & 0.013 \nl
without tidal dwarfs (32)        & 4.2 &  $6.2 \pm 1.3$ & $1.5 \pm 0.3$
 & \nl
 & & & & \nl
with 1st-ranked E/S0 (14) & 4.7 & $11.5 \pm 2.4$ & $2.4 \pm 0.5$
 & 0.012 \nl
with 1st-ranked S (25)    & 3.9 &  $3.7 \pm 1.3$ & $0.9 \pm 0.3$ & \nl
 & & & & \nl
with x-ray emission (7)      & 4.6 & $15.0 \pm 2.8$ & $3.3 \pm 0.6$
 & $< 0.001$ \nl
without x-ray emission (32)  & 4.1 &  $3.4 \pm 1.2$ & $0.8 \pm 0.3$ & \nl
 & & & & \nl
$M/L < 50h^{\ast \ast}$ (21)      & 3.9 & $5.7 \pm 1.3$ & $1.5 \pm 0.3$
 & 0.535 \nl
$M/L \ge 50h$ (16)                & 4.6 & $7.1 \pm 2.3$ & $1.5 \pm 0.5$
 & \nl
 & & & & \nl
velocity dispersion $< 200{\rm km/s}^{\ast \ast}$ (20)  & 3.8 &
 $5.4 \pm 1.4$ & $1.4 \pm 0.4$ & 0.744 \nl
velocity dispersion $> 200{\rm km/s}$ (19)              & 4.5 &
 $6.7 \pm 1.9$ & $1.5 \pm 0.4$ & \nl
 & & & & \nl
spiral fraction $<50\%$ (19)    & 4.6 & $7.7 \pm 2.1$ & $1.7 \pm 0.5$
 & 0.729 \nl
spiral fraction $\ge 50\%$ (20) & 3.8 & $5.2 \pm 1.4$ & $1.4 \pm 0.4$
 & \nl
 & & & & \nl
median separation $< 39h^{-1}{\rm kpc}^{\ast \ast}$ (20) & 4.1
 & $5.4 \pm 1.3$ & $1.3 \pm 0.3$ &0.355 \nl
median separation $> 39h^{-1}{\rm kpc}$ (19)             & 4.3
 & $6.5 \pm 1.9$ & $1.5 \pm 0.4$ & \nl
 & & & & \nl
\# giant members $\le$ 4 (26)                     & 3.6 &  $3.5 \pm 1.3$
 & $1.0 \pm 0.4$ & 0.004 \nl
\# giant members $>$ 4 (13$^{\dagger \dagger}$)   & 5.3 & $10.8 \pm 2.1$
 & $2.0 \pm 0.4$ & \nl
 & & & & \nl
groups without x-ray emission &&&&\nl
\hspace*{2em}$M/L < 50h$ (18)   & 3.9 & $ 5.1 \pm 1.4$ & $ 1.3 \pm 0.4$
 & 0.028 \nl
\hspace*{2em}$M/L \ge 50h$ (12) & 4.5 & $-3.9 \pm 2.7$ & $-0.9 \pm 0.6$ \nl
 & & & & \nl
\enddata
\tablenotetext{} {$^\ast$ Dwarf galaxy counts include objects in the
luminosity range $-18.0 < M_R < -14.0$ within $R=1.50 R_G$ unless otherwise
stated.}
\tablenotetext{} {$^\dagger$ Calculated values are based on a comparison of
faint end data points of the appropriate luminosity functions.
A value of $P_{\chi}$ approaching zero implies the two data sets are from
different populations.}
\tablenotetext{} {$^\ddagger$ Tidal dwarf candidates are included
only in these subsets.}
\tablenotetext{} {$^{\ast \ast}$ This is the median value of the Hickson
catalog.}
\tablenotetext{} {$^{\dagger \dagger}$ We note that 8 of the 13 groups
have dominant E/S0 galaxies and/or x-ray halos.}
\end{deluxetable}

\end{document}